\documentstyle[twocolumn,prl,aps,psfig]{revtex}
\begin{document}
\draft
\narrowtext
\title{Strained tetragonal states and Bain paths in metals}
\author{P. Alippi$^1$, P.M. Marcus$^{1,2}$, and M. Scheffler$^1$}
\address{${}^1$Fritz-Haber-Institut der Max-Planck-Gesellschaft, 
Faradayweg 4-6, D-14195 Berlin-Dahlem, Germany \\
${}^2$IBM Research Center, Yorktown Heights, N.Y. 10598, USA}
\date{\today}
\twocolumn[
\maketitle
\vspace{-0.5cm}
\begin{quote}
\parbox{16cm}{\small
Paths of tetragonal states between two phases of a material, 
such as bcc and fcc, are called Bain paths. Two simple Bain 
paths can be defined in terms of special imposed stresses, one of
which applies directly to strained epitaxial films. Each path goes 
far into the range of nonlinear elasticity and reaches a range of 
structural parameters in which the structure is inherently 
unstable. In this paper we identify and analyze the general properties
of these paths by density functional theory. Special examples 
include vanadium, cobalt and copper, and the epitaxial path is used to
identify an epitaxial film as related uniquely to a bulk phase.\\

}
\end{quote}]

Pseudomorphic epitaxy of a cubic or tetragonal (001) film 
typically results in a strained tetragonal structure. 
If the stresses on  the tetragonal state vanish and also 
the state corresponds to a local minimum of energy with 
respect to tetragonal deformations, the structure will 
be called a {\em {tetragonal phase}}. 
Such a phase will be stable or metastable depending
on whether it has the lowest energy compared to other minima.  
Frequently metals have two tetragonal phases; sometimes
both are cubic, e.g., bcc and fcc Na and Rb~\cite{Milst_old,Mohn}; 
sometimes one is cubic and the other phase is tetragonal, 
e.g., Ti and V~\cite{Mohn,Craiev_prl}. They can, of course, 
also have phases with other structures, e.g., Ti also has a hcp 
phase. 

Many paths can go from one tetragonal phase to the other. 
If the geometries along such a path have tetragonal symmetry and if
they connect bcc and fcc phases, the paths have been called 
Bain paths~\cite{Bain}. 
A purpose of the present work is to define and discuss a 
particular Bain path which will be called the epitaxial 
Bain path (EBP). The EBP is produced by isotropic stress 
or strain in the (001) plane of tetragonal phases 
accompanied by vanishing stress perpendicular to the 
plane, such as pseudomorphic epitaxy produces on an (001) 
cubic or tetragonal film.
Epitaxy provides a valuable means of stabilizing metastable 
phases and of putting phases under very large strains, 
both tensile and compressive in the plane of the epitaxy. 
The EBP of a material identifies the phase that has been 
strained, checks quantitatively the elastic behavior, 
which can be highly nonlinear, and predicts which phase 
of the material will form on a given substrate.  Thus, in 
order to understand the properties of epitaxial films and new
materials, the knowledge of the EBP is indispensable.

A different Bain path has long been discussed, particularly 
by Milstein~\cite{Milst_old}, in which uniaxial stress is applied to 
a tetragonal state along the [001] axis accompanied by 
zero stress in the (001) plane; this path is conveniently 
called the uniaxial Bain path (UBP). We compare the 
two paths, EBP and UBP, which are both physically 
realizable.  We show that both have the same lowest possible 
maximum energy or barrier energy of all Bain paths between 
the two tetragonal phases. However the EBP has a special 
value in relating strained tetragonal structures 
produced by epitaxy to particular tetragonal phases.
Depending on where the tetragonal structure lies 
on the EBP, we shall show that a clear choice can be made 
which classifies the structure as a strained form 
of a particular phase. 
Thus for materials with both a bcc and a fcc phase, the bulk (i.e., 
the interior) of an epitaxial film can be uniquely classified 
as a strained bcc phase or a strained fcc phase. 
Experimental geometry analyses of the bulk structure
are frequently made by low-energy electron diffraction
of ultrathin epitaxial films, which can then be placed on the 
EBP in the tetragonal plane to identify the phase of which it is the
strained form.

We  employ density functional theory together with the 
full-potential linearized augmented plane wave 
method~\cite{WIEN}.  An important advantage of a 
first-principles calculation, i.e., a calculation with no 
adjustable parameters, is that it applies just as reliably 
to highly strained or even unstable geometries, as it does 
to the unconstrained ground state, i.e., to the stable 
phase. The same reliability cannot be 
asserted for empirically adjusted potentials, such as 
the empirical pseudopotential model used in Ref.\cite{Milst_old}, 
which is adjusted to reproduce ground-state properties of 
only certain structures.
The first-principles calculation provides a check on the 
validity of empirical potentials far from the structural 
ground state and of linear elasticity theory. 

\begin{figure}
\centerline{\hbox{\hspace{-1cm}
\psfig{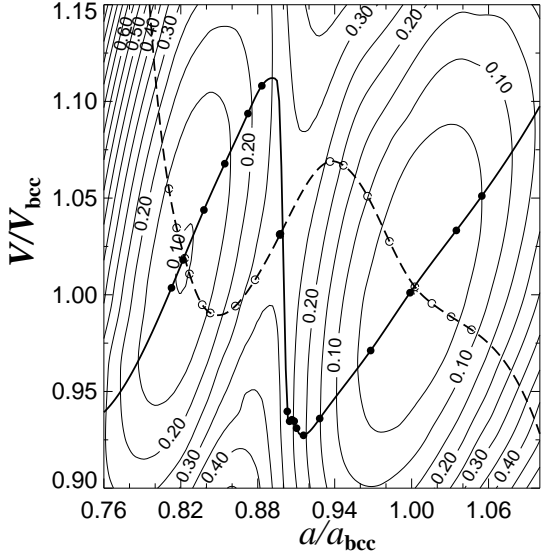}}}
\caption{Contour lines of constant $E$ in eV/atom for vanadium 
on the tetragonal plane showing a bcc minimum 
(the zero of $E$) with $\protect{a_{\rm bcc}=2.93\,{\mbox{\AA}}}$ and 
$\protect{V_{\rm bcc}=12.6\,{\mbox{\AA}^3}}$ ($\protect{c/a=1}$), 
a minimum at $\protect{a=2.41\,{\mbox{\AA}}} \, (c/a=1.83)$ 
and $\protect{V=12.8\,{\mbox{\AA}^3}}$, 
and a saddle point at 
$\protect{a=2.65\,{\mbox{\AA}}} \, (c/a={(2)^{\frac{1}{2}}})$ and 
$\protect{V=13.1\,{\mbox{\AA}}^3}$. 
The epitaxial Bain path (EBP, full dots) 
and the uniaxial Bain path (UBP, open dots) are also shown.}
\label{v_aV_cont}
\end{figure}

The calculation proceeds by evaluating the total energy 
$E(a,c)$ in eV/atom as a function of the tetragonal 
lattice constants $a$ and $c$.  Contour lines of constant 
$E$ are plotted in Fig.\ref{v_aV_cont} for vanadium on the 
($a/a_{\rm bcc}\,$-$\,V/V_{\rm bcc}$) plane, where $V=ca^2/2$ is the 
volume per atom. 
The contour lines show clearly two minima, one at the 
bcc structure and one at a tetragonal structure, and 
between them a saddle point at the fcc structure.  
Thus vanadium has two fairly deep and well-separated 
minima and so is a good choice to illustrate the 
tetragonal paths between minima~\cite{note_mohn}. 
Since $E$ has an extremum at the cubic 
structures~\cite{Craiev_prl,Kraft_prb}, the saddle point 
will occur exactly at the fcc structure. 

Also plotted in Fig.\ref{v_aV_cont} are the EBP and the UBP. The 
EBP is calculated from the minima of $E(c)$ at constant 
$a$, which corresponds to the epitaxial situation, i.e., 
$a$ is held fixed and $c$ adjusts to minimize $E(c)$ and 
thus makes the out-of-plane stress, i.e., the  stress 
along [001], vanish. Conversely, the UBP calculation fixes $c$ 
and lets $a$ adjust to minimize $E(a)$ making
the in-plane 
stress vanish. Both paths go through the minima and the 
saddle point, hence the maximum energy on each path is 
the same. 

The saddle point is called in Ref.\cite{Milst_old} ``a special 
unstressed tetragonal state ... at a local energy 
maximum''~\cite{note2}. Obviously any path on the   
tetragonal plane that does not go through the saddle 
point must go through states of higher energy than the 
saddle point, since it will cross contours in either the 
upper or lower central sectors formed by the saddle-point 
contours at $E=0.29$ eV/atom. 
Two common paths that do not go through the saddle point, 
hence have higher maximum energies than the EBP or UBP, 
are : 1. the constant volume paths at the volumes of the 
minima, which differ from the saddle-point volume. These 
paths have been called volume conserving Bain 
paths~\cite{Mohn,Craiev_prl}; 2. the hydrostatic 
pressure path, which has equal in-plane and out-of-plane 
stress components.

\begin{figure}
\centerline{\hbox{
\psfig{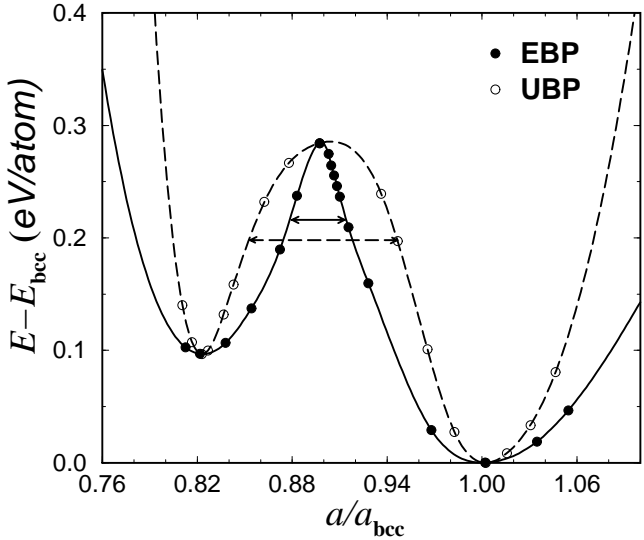}}}
\caption{Plots of $E$ in eV/atom along the Bain paths EBP and 
UBP for vanadium between the minima and through the saddle 
point. The unstable ranges of $a$ are marked with arrows 
for both paths.}
\label{bain_Evsa}
\end{figure}

Figure~\ref{bain_Evsa} plots $E$ vs. $a$ along the EBP and UBP.
Also indicated is the range from $a=2.58 \, \mbox{\AA}\, (c/a=1.6)$ 
to $a=2.66 \, \mbox{\AA}\, (c/a=1.2)$ for which we  find along the EPB
the tetragonal structure to be inherently unstable~\cite{note3}. In this
range the slope of the EBP changes sign, and
the stability condition 
\begin{eqnarray}
\tilde c_{11} \tilde c_{33} - 
\tilde c_{13}^2 \geq 0 
\label{eq1}
\end{eqnarray}
is violated. Here  $\tilde c_{ij}$ are calculated from 
$\tilde c_{11} =(a^2 /V)\partial^2 E(a,c)/{\partial a}^2$, 
$\tilde c_{13} =(ac /V)\partial^2 E(a,c)/\partial a \partial c$, 
$\tilde c_{33} =(c^2 /V)\partial^2 E(a,c)/{\partial c}^2$ 
for the strained lattice~\cite{note_inst}. 
Tetragonal states outside the range of Eq.(\ref{eq1}) cannot 
be stabilized with any applied stresses. When a path on the 
tetragonal plane reaches the boundaries of this unstable 
region, the tetragonal crystal becomes unstable and an 
epitaxial film will not be able to form many layers at the 
bulk spacing. Epitaxial films strained to a state near the 
unstable region should show large changes in the phonon spectra 
as elastic constants weaken, which would be interesting to study.

One consequence of the existence of an unstable region is 
that a clear conceptual distinction can be made between 
strained states of the phase at lower $c/a$ (for vanadium 
the bcc phase) and strained states of the larger $c/a$ phase 
(for vanadium at $c/a=1.83$). If there was a continuous 
path of allowed states between the two phases, there would 
not be a clear criterion for regarding an observed strained 
state as strained from a particular one of the two phases. 

A second consequence of the unstable region is that 
the strained system never gets to the saddle point, 
but transforms to another strained phase at an energy at 
or below the energy of the state where the path strikes 
the boundary of the unstable region. Thus for the EBP and 
UBP the breakdown energy is at least 0.05 eV/atom lower than
the saddle-point energy, and it is lower on the UBP.

\begin{figure}
\centerline{\hbox{
\psfig{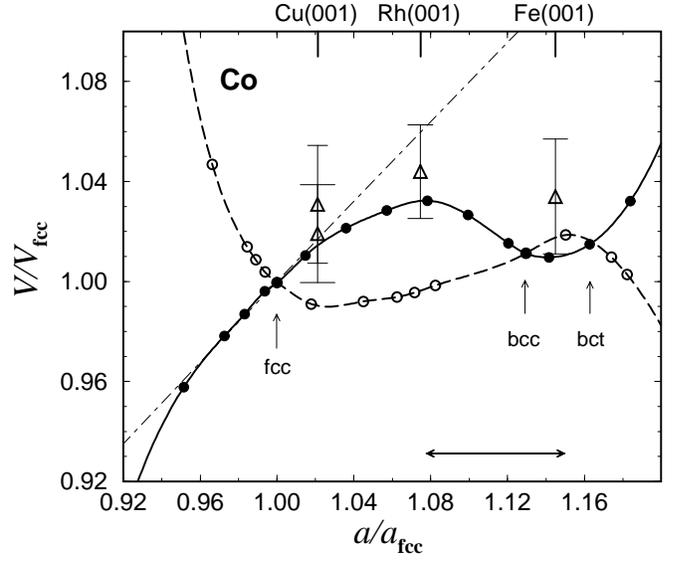}}}
\caption{Epitaxial Bain paths on the ($a/a_{\rm fcc}$-$V/V_{\rm fcc}$) 
plane for Co, with LEED bulk structures for strained 
Co/Rh(001)\protect\cite{Co_Rh}, Co/Cu(001)\protect\cite{Co_Cu}, 
Co/Fe(001)\protect\cite{Co_Fe}. The error limits are shown 
by a line through the point and come entirely from the 
estimated error in $c$ from experimental analysis. 
The theoretical value of $a_{\rm fcc}$ for the fcc phase 
is $2.50 \, \mbox{\AA}$ ($\protect{V_{\rm fcc} = 11.09 \,
{\mbox{\AA}}^3}$), which agrees with the experimental 
value\protect\cite{exper}. The dot-dashed lines are obtained from 
linear elasticity theory, with elastic constants of the fcc 
structure. The horizontal arrow mark the range along the EBP 
path where Eq.(\protect\ref{eq1}) is violated.}
\label{Co}
\end{figure}

\begin{figure}
\centerline{\hbox{
\psfig{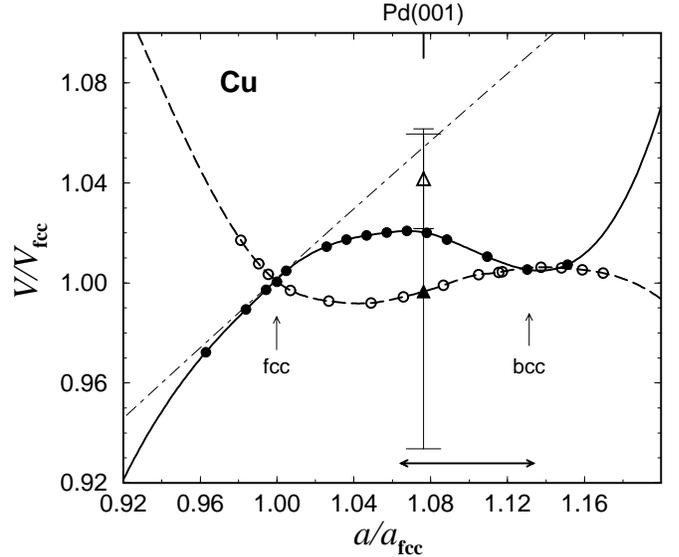}}}
\caption{Same as previous figure, but for Cu. The 
LEED\protect\cite{Cu_Pd} (open triangle) and the
STM\protect\cite{hahn_prl} (full triangle) bulk results for Cu/Pd(001)
are given as well. The theoretical value of $a_{\rm fcc}$ for Cu is
$2.53 \, \mbox{\AA}$ ($V_{\rm fcc} = 11.52 \, {\mbox{\AA}^3}$), while
the experimental points use the experimental value of 
$a_{\rm fcc} = 2.56 \, \mbox{\AA}$ ($V_{\rm fcc} = 11.76 \, {\mbox{\AA}^3}$).}
\label{Cu}
\end{figure}

The EBP and UBP can be found for other metals by similar 
calculations; for Cu and Co our results are shown in Fig.\ref{Co} and
Fig.\ref{Cu}. 
The topology of the $E(a,c)$ surface of cobalt is similar 
to the one of vanadium, with the two minima and the saddle 
point corresponding to the structures where the EBP and the 
UBP intersect. In this case, however, the stable cubic phase 
is face-centered ($c/a=\sqrt2$) with 
$a_{\rm fcc} = 2.50 \, \mbox{\AA}$, the bcc structure 
corresponds to the saddle point ($a_{\rm bcc}=2.83\, \mbox{\AA}$) 
and a metastable tetragonal structure is found at 
$c/a = 0.92$ ($a=2.91 \, \mbox{\AA}$)~\cite{Liu}.
The energy barrier $\protect{E_{\rm bcc}-E_{\rm fcc}}$ is $72 \, 
\mbox{meV}$.

The situation is different for Cu: the stable phase is at
the fcc structure ($a_{\rm fcc}=2.53 \, \mbox{\AA}$), but 
there is no other tetragonal metastable phase, and the 
energy surface $E(a,c)$ shows a large plateau near the bcc 
structure~\cite{jeong}; for this reason, the EBP and the UBP 
share a common part for $1.13 \leq a/a_{\rm fcc} \leq 1.15$.

Strained structures found by LEED are entered in the graphs; 
in each case the error limits intersect the EBP. 
Prediction of which phase will be produced and how strained 
it will be by a given substrate can be immediately made by 
entering the $a$ of the substrate on the $a$ axis of the EBP.

For Co on Rh(001)~\cite{Co_Rh} and on Cu(001)~\cite{Co_Cu} the 
structures lie on the left of the unstable region and thus both 
correspond to strained fcc structures. However the error line 
for Co on Fe(001)~\cite{Co_Fe} is at a value of $a$ above the 
unstable range and must be considered a strained form of the 
tetragonal phase of Co at $c/a=0.92$; bcc Co itself at $c/a=1$ 
($a=2.83$ \AA) is unstable.

The strained tetragonal structure for Cu on Pd$(001)$
lies in the range of the EBP identified as unstable by our theory.
We therefore disagree with the conclusions of a recent
scanning tunneling microscopy (STM) experiment~\cite{hahn_prl}, 
where such a film was ``identified'' as a metastable tetragonal phase. 
The structure measured in the high coverage regime ($1\!-\!7$ layers
at T$=300$ K, $3\!-\!7$ layers at higher temperatures) agreed within 
experimental error with the LEED analysis~\cite{Cu_Pd}; however, the 
morphology of the islands observed in the STM pictures 
was taken as indication of a structure under small planar 
strain, and, therefore, of the existence of a Cu tetragonal 
phase with lattice constant close to $a_{Pd}$. 
Our results predict that there is no tetragonal phase at 
$c/a < \sqrt{2}$ (and $a > a_{\rm fcc}$); we also show 
that the film is too far from the bcc region of the EBP to 
be regarded as strained bcc. Thus, we conclude that the properties of 
such Cu/Pd(001) film are not bulk like, but instead stabilized by 
the (still ultrathin) film properties.

In Fig.\ref{Cu} (as in Fig.\ref{Co}) the epitaxial paths from 
linear elasticity theory are also drawn for fcc Cu (and Co). 
Epitaxial strain on a cubic $(001)$ surface in the linear elastic 
approximation says normal stress vanishes, hence 
$\delta c / c = - (2\nu / (1-\nu)) \, \delta a / a$, 
where $\nu$ is the Poisson ratio. A comparison with the 
calculated EBP in both figures shows how the linear 
approximation fails to be valid for $\delta a / a$ 
equals to few percent. In the analysis of the 
Cu/Pd$(100)$ film, it is evident that the film is already 
far from the linear elastic regime. 

Generalizations of this calculation of physically 
realizable paths between phases are possible with existing 
codes. Milstein~\cite{Milst_lett} has considered the 
instability of tetragonal structures with respect to 
orthorhombic deformations and found such instabilities on 
the UBP outside the range of $c/a$ between the two minima; 
these instabilities probably exist on the EBP also, but 
have not as yet been studied. It looks practicable to 
explore with present codes even more general paths of 
homogeneous strain between tetragonal phases that make 
more general deformations of the unit mesh than 
orthorhombic deformations. 
Possibly a path of deformed lattices could be found with
lower maximum energies, as the path of trigonal lattices 
of Ref.~\cite{Kraft_prb}. 
It would also be possible to find epitaxial paths of 
tetragonal states for slabs of a finite number of layers, 
which would include relaxations of surface layers. 

In summary, paths of tetragonal states between two 
tetragonal phases include two particularly simple cases, 
namely, those produced by epitaxial (isotropic biaxial) 
stress and by uniaxial stress.
Both cases correspond to maintaining certain surfaces 
stress-free and both have minimal maximum energies at 
a saddle point of energy that both pass through. 
Along these paths a range of instability exists, which 
separates states that can be clearly regarded as 
strained from one or the other of the two phases. 
The epitaxial path permits definite identification 
of the equilibrium phases of strained epitaxial films 
and could usefully be found for all metals with 
tetragonal phases.

\begin{references}
\bibitem{note_mohn}
A plot of energy contour lines for vanadium  is also given 
in Ref.2, which provided in addition a useful power series 
expansion of $E$ with which the present calculations were checked.
\bibitem{note2} 
The saddle point is not actually a local maximum of energy, since 
$E$ increases from that point in the central sectors formed 
by the saddle-point contours at $E=0.29$ eV/atom in 
Fig.~\ref{v_aV_cont}.
\bibitem{note3} 
At all $V$ in Fig.~\ref{v_aV_cont}, the unstable region 
boundaries have about the same $c/a$ values.
\end{references}
\end{document}